\documentclass[aps,twocolumn,superscriptaddress,groupedaddress,10pt,nofootinbib]{revtex4} 

\usepackage{graphicx}  
\usepackage{dcolumn}   
\usepackage{bm}        
\usepackage{amssymb}   
\usepackage{amsmath}
\usepackage{epstopdf}
\usepackage[a4paper,margin=20mm]{geometry}
\usepackage{color}

\begin{document}

\title{Constitutive relations for shear fronts in shear-thickening suspensions}

\author{Endao Han}
\email[E-mail: ]{endao.han1988@gmail.com}
\affiliation{James Franck Institute and Department of Physics, The University of Chicago, Chicago, Illinois 60637, USA}
\author{Matthieu Wyart}
\affiliation{Institute of Physics, \'Ecole Polytechnique F\'ed\'erale de Lausanne, CH-1015 Lausanne, Switzerland}
\author{Ivo R. Peters}
\affiliation{Engineering and the Environment, University of Southampton, Highfield, Southampton SO17 1BJ, UK}
\author{Heinrich M. Jaeger} 
\affiliation{James Franck Institute and Department of Physics, The University of Chicago, Chicago, Illinois 60637, USA}

\date{\today}

\begin{abstract}
We study the fronts that appear when a shear-thickening suspension is submitted to a sudden driving force at a boundary. Using a quasi-one-dimensional experimental geometry, we extract the front shape and the propagation speed from the suspension flow field and map out their dependence on applied shear. We find that the relation between stress and velocity is quadratic, as is generally true for inertial effects in liquids, but with a pre-factor that can be much larger than the material density. We show that these experimental findings can be explained by an extension of the Wyart-Cates model, which was originally developed to describe steady-state shear-thickening. This is achieved by introducing a sole additional parameter: the characteristic strain scale that controls the crossover from start-up response to steady-state behavior. The theoretical framework we obtain unifies both transient and steady-state properties of shear-thickening materials. 
\end{abstract}

\maketitle



Dense suspensions of micron-sized solid particles in a simple liquid can show a rich set of properties under an imposed shear stress, such as continuous shear thickening (CST) \cite{Wagner_book, Brady}, discontinuous shear thickening (DST) \cite{Barnes_DST, DST_review, DST_JOR} and jamming \cite{Ivo_Nature}. DST is a striking phenomenon whereby the viscosity of the suspension shoots up discontinuously when a certain shear rate threshold is reached. Jamming is an even more extreme case where the suspension becomes not just more viscous but turns into a solid with non-zero shear modulus. These strongly non-Newtonian rheological properties become most pronounced at high particle packing fractions, where suspensions start to exhibit characteristics also found in dry granular material \cite{Cates_Jamming, OHern, SJ, DeGiuli15a, Lerner12a}.
In particular, recent experimental \cite{MicroMechST, Boyer, Reverse_Itai, Royer, Colin_2017, Forterre_2017} and numerical work \cite{Seto, Mari, Ness, Morris2017}  points to the existence of a stress threshold above which the dominant interaction between particles switches from hydrodynamic frictionless lubrication  to granular friction forces. This crossover forms the basis of a phenomenological model developed by Wyart and Cates, which unifies CST, DST and jamming under a common framework \cite{Wyart_Cates}.

Efforts to map out a state diagram that delineates the properties of dense suspensions as a function of packing fraction and imposed forcing have focused almost exclusively on steady-state conditions.  This does not capture the many remarkable transient phenomena exhibited by suspensions \cite{vonKann, Zhang, Scott, Ivo_2D, Stone, Endao, Smith_NC, Sayantan, Ivo_Nature}. Among these, impact-activated solidification is commonly referred to, but also is one of the least well understood. Only a couple years ago was it discovered \cite{Scott, Ivo_2D} that this solidification is not simply strong shear thickening, as previously assumed, but a dynamic process where impact at the suspension surface initiates jamming fronts that rapidly propagate into the bulk of the material. Recent ultrasound experiments revealed that these fronts constitute localized bands of intense shear that transforms the suspension from a fluid-like into a solid-like state without measurably increasing the particle packing fraction \cite{Endao, Sayantan}. So far, however, several key aspects have remained largely unresolved, including (i) the conditions under which dense suspensions can develop jamming fronts; (ii) the shape of the flow profile at the front; and importantly (iii) the constitutive relation between the applied stress and the front speed. These questions underline the need to build a description that would encompass both transient and steady-state properties of shear thickening materials.

In this Letter, we consider the arguably simplest geometry in which these properties can be measured: a plane of fluid that is sheared along one of its boundaries. With a resulting flow field that changes only along the direction perpendicular to the sheared boundary, this configuration exhibits one-dimensional (1D) dynamics. A key finding is that the velocity-stress relation measured at the boundary, i.e., the macroscopic response of the suspension to applied forcing, is governed by a microscopic, particle-scale quantity: the amount of strain accumulated locally when the jamming front passes through. This accumulated strain depends on both the intrinsic properties of the suspension, such as the packing fraction, and the system's initial preparation condition.  We can capture this transient behavior with the Wyart-Cates model by introducing one additional parameter, a strain amplitude $\gamma^*$ characterizing the cross-over to steady-state flow. With this generalization the model exhibits well-defined jamming fronts and allows us to compute their dependence on packing fraction and forcing conditions, leading to predictions for (i,ii,iii) in close agreement with our experimental observations. 


\subsection*{Quasi-one dimensional shear experiments}
\label{sec:FrontDemo_EXPT}

\begin{figure*}[t!]
\begin{center}
\includegraphics[scale=0.85]{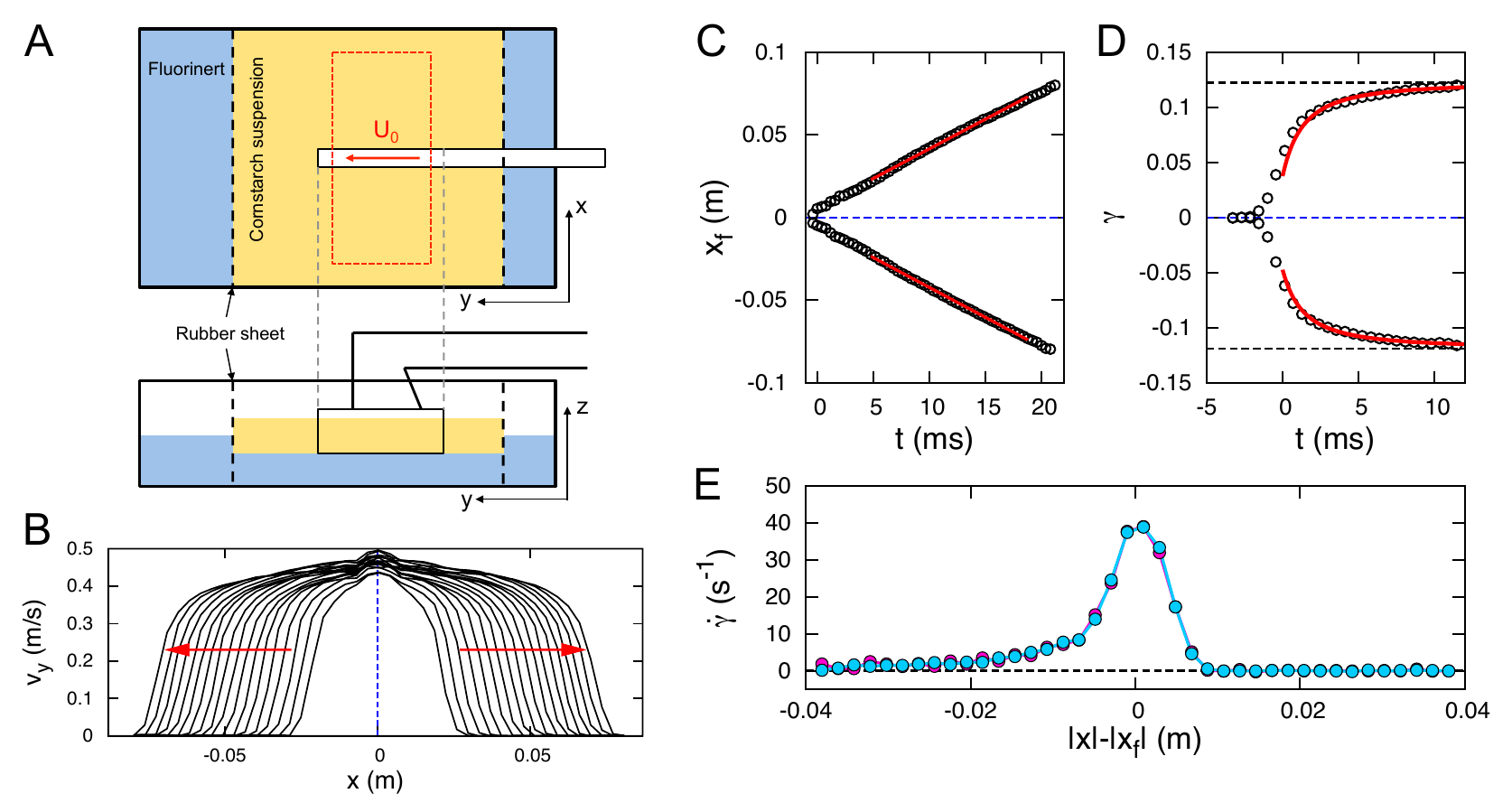}
\end{center}
\caption{Quasi-1D shear experiment. (A) Illustration of the experimental setup, consisting of a layer of cornstarch suspension (yellow) that floats on oil (blue). Dashed black lines represent rubber sheets confining the suspension. An acrylic plate with roughened surface was inserted in the middle of the container (at $x = 0$~m) and moved with speed $U_0$. The dashed red box indicates the area used for data analysis. (B) Exemplary velocity profiles of a shear front for $\phi = 0.532$ and $U_0 = 0.46 \pm 0.02$~m/s, propagating transversely to either side of the plate (dashed blue line). (C, D) Front position $x_\text{f}$ and accumulated strain $\gamma$ as functions of time $t$.  In C, red lines show linear fits. In D, time $t = 0$~ms represents the time when $x = x_\text{f}$. Red lines show fits to a power law. Black dashed lines indicate the asymptotic accumulated strain $\gamma_\infty$. (E) Local shear rate calculated from the mean velocity profiles. Blue and pink represent the left and right branches, respectively, in B.}
\label{Front_EXPT}
\end{figure*}

Our experimental system, based on Ref~\cite{Ivo_2D} and illustrated schematically in Fig.~\ref{Front_EXPT}, consisted of a layer of cornstarch suspension into which a thin plate was inserted. Starting from rest, the plate was impacted by a computer-controlled linear actuator (Parker ETT050) and then moved along the y-direction at constant speed $U_0$. The suspension was floated on heavy, low-viscosity oil (Fluorinert FC-3283 from 3M), providing a nearly stress-free boundary condition. This allowed us to deduce the stress applied at the boundary from the momentum of the suspensions, which we measured experimentally. A high-speed camera (Phantom V12) was used to image the motion of the suspension surface. The videos were analyzed using a particle imaging velocimetry (PIV) algorithm to obtain the flow field.


Inside the dashed red box in Fig.~\ref{Front_EXPT}A, the system is, to very good approximation, quasi-1D, with significant flow field gradients only along the $x$ direction. 
We therefore average in the $y$ direction and leave $x$ as the only spatial coordinate. An example of a flow field exhibiting a front is shown in Supplementary Movie 1. Fig.~\ref{Front_EXPT}B shows the evolution of the resulting, averaged velocity profiles. As the arrows indicate, a moving region rapidly expands outward to either side of the plate, while the shape of the velocity profiles stays approximately invariant.

For convenience, we define the front position $x_\text{f}$ as the point on a velocity profile where $v_y = 0.45 U_0$. As shown in Fig.~\ref{Front_EXPT}C, $x_\text{f}$ is a linear function of time on both sides of the plate, providing a well-defined, constant front propagation speed $U_\text{f}$. In the example shown $U_\text{f} = 3.60 \pm 0.03$~m/s, which is 7.8 times the plate speed $U_0 = 0.46 \pm 0.02$~m/s, but much slower than the speed of sound in the material (about 1900~m/s \cite{Endao}).
From the flow fields, we also extract the local accumulated strain $\gamma$ as a function of time. Because of the invariance of the velocity profiles, we collapse the $\gamma$-$t$ curves at different $x$ using the time when the front reaches that position, \textit{i.e.} $x_\text{f}(t) = x$. As shown in Fig.~\ref{Front_EXPT}D, $\gamma$ increases quickly at the beginning, but then slows down because of shear thickening. The red curves are power law fits to the data for $t > 0$~ms and indicate that the accumulated strain will approach  $0.12$ asymptotically. This asymptotic approach to a finite value of the accumulated strain (under continued finite stress) is a clear indicator of jamming

To obtain the shear rate distribution along the velocity profiles we average them after shifting the front positions $x_\text{f}$ to zero. The absolute value of the local shear rate $|\dot{\gamma}| = |d v_y / d x|$ is shown in Fig.~\ref{Front_EXPT}E \footnote[1]{Since in our 1D system, $\gamma$ always changes monotonically, we only consider the absolute value of $\dot{\gamma}$ and $\gamma$. In the paper the signs on $\dot{\gamma}$ and $\gamma$ are ignored.}. The maximum shear rate $\dot{\gamma}_\text{max}$ is found close to the front position. However, the shear rate profile is not symmetric with respect to $|x|-|x_\text{f}| = 0$, exhibiting a steeper gradient at the leading edge ($|x| > |x_\text{f}|$). Behind the front, we observe a tail of small, but finite shear rate. Thus, strictly speaking, the region in the wake of the passing front does not immediately become solid-like. However, as the front keeps moving ahead and the local strain approaches its finite asymptotic value $\gamma_\infty$, a jammed state with non-zero yield stress is reached. By contrast, if a suspension does not ``jam'' but only shear ``thickens'', we expect the accumulated strain to keep growing and the shear rate to stay finite.

From this discussion, we extract three defining features for jamming fronts:
1.~A well-defined, step-like velocity profile that stays invariant over time;
2.~A constant propagation speed $U_\text{f}$;
3.~An asymptotically accumulated strain that stays finite.
These characteristics distinguish jamming fronts from the more diffusive response to applied shear that occurs at low driving speeds $U_0$ or at low packing fractions, as discussed below. We will use them in comparing model calculations to the experiments.


\begin{figure*}[!t]
\centering
\includegraphics[scale = 0.9]{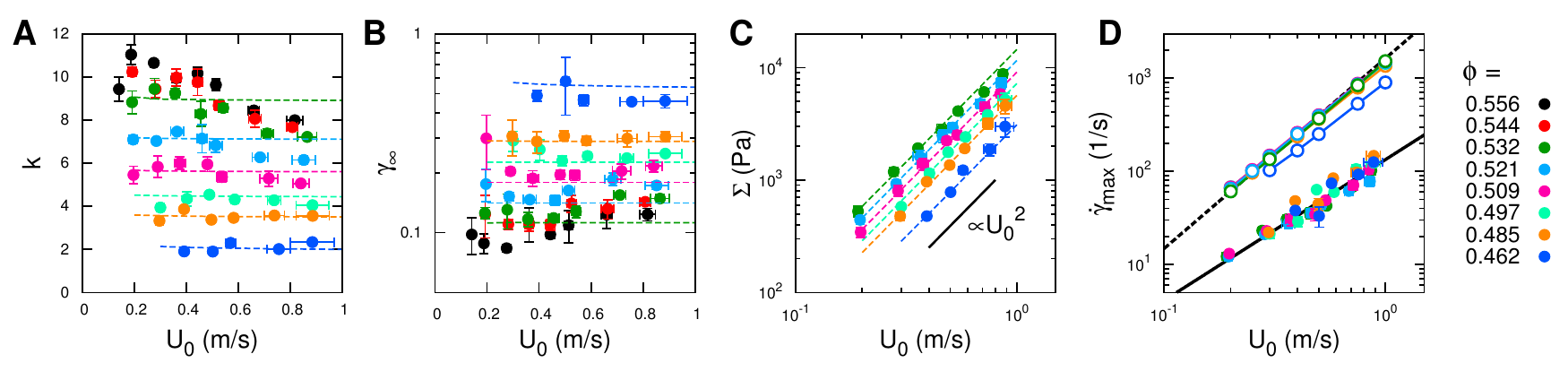}
\caption{Characteristics of propagating jamming fronts as function of pushing speed $U_0$. Experimentally obtained data for different packing fractions $\phi$ are shown by solid symbols. In (A-C), dashed lines are from model calculations. (A) Dimensionless front propagation speed $k$. (B) Asymptotic accumulated strain $\gamma_\infty$. (C) Stress at the boundary $\Sigma$. Solid circles are calculated by plugging experimentally measured $U_0$ and $k$ into Eq.~\ref{eq:Stress}. Dashed lines show the stress at the boundary obtained from the numerical calculations, which satisfy $\Sigma \sim U_0^2$. (D) Maximum shear rate $\dot{\gamma}_\text{max}$. Open circles are from the model (the same color scheme is used as for data from the experiments). Black lines are power law fits.}
\label{Compare}
\end{figure*}

We performed the quasi-1D shear experiments at different packing fractions $\phi$ and pushing speeds $U_0$. At slow $U_0$, the suspension is fluid like, and we obtain $U_\text{f} \approx 0$~m/s (see Supplementary Information). As $U_0$ increases, we start to see a front that propagates out. However, before $U_0$ is sufficiently fast, the flow does not have the three features of the jamming fronts. For example, there is no well defined $U_\text{f}$ at intermediate driving speed. We now focus on the limit of sufficiently fast $U_0$, where we obtain jamming fronts as defined above.


\subsection*{Front speed and accumulated strain} 
\begin{figure}[t!]
\begin{center}
\includegraphics[scale=1]{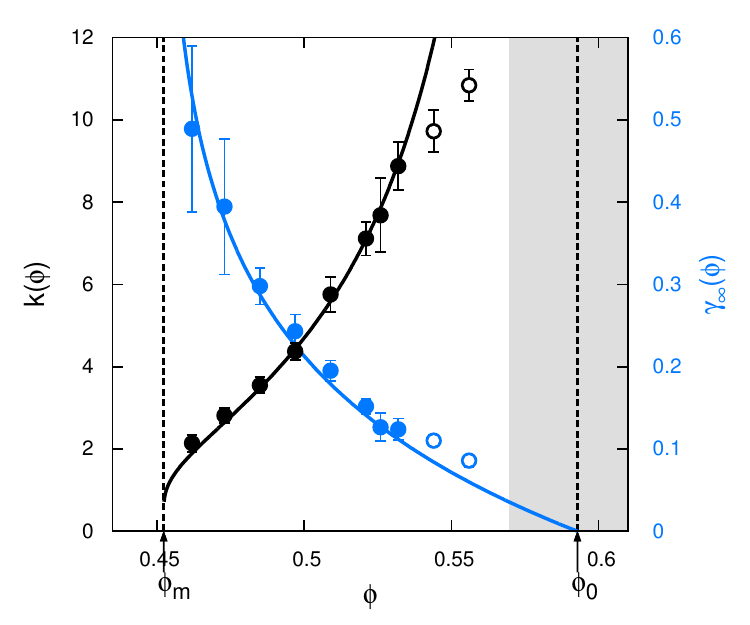}
\end{center}
\caption{\label{K_SS_phi} Plateau of dimensionless front propagation speed $k_\text{p}$ (black) and asymptotic accumulated strain $\gamma_\text{p}$ (blue) as functions of packing fraction $\phi$. The blue curve shows Eq.~\ref{eq:FP}, and the black curve shows its reciprocal, both with $\gamma^* = 0.197$. Data at $\phi = 0.556$ and 0.544 are represented by open circles. }
\end{figure}

Here we consider how $U_\text{f}$, $\gamma_\infty$, the stress at the boundary $\Sigma$, and the maximum shear rate $\dot{\gamma}_\text{max}$ depend on the wall velocity $U_0$ at different $\phi$, as shown in Fig.~\ref{Compare}. 
We define the normalized front propagation speed as $ k \equiv U_\text{f} / U_0 $. The variation of $k$ as a function of $U_0$ is presented in Fig.~\ref{Compare}A, and Fig.~\ref{Compare}B shows the corresponding $\gamma_\infty$.
At $\phi \leq 0.5$, both $k$ and $\gamma_\infty$ are essentially constant. However, for the largest values of $\phi$ and $U_0$ we probed, departure from this constant behavior can be detected, an effect whose relative magnitude can  be as large as $30\%$. 
Still, for each $\phi < 0.54$, we can find a range in which $k$ and $\gamma_\infty$ are nearly independent of $U_0$. Using the average value in such a range, we can define $k(\phi)$ and $\gamma_\infty(\phi)$ at each $\phi$. Note that $\gamma_\infty (\phi)$ decreases with increasing $\phi$, and the trend is reversed for $k(\phi)$. These results are documented in Fig.\ref{K_SS_phi}. 

Visual inspection of Fig.~\ref{Compare}A, Fig.~\ref{Compare}B and Fig.\ref{K_SS_phi} suggests an inverse relationship between $k$ and $\gamma_\infty$, which we now derive. 
The total accumulated strain when the front passes through is
$\gamma_\infty = \int^{+\infty}_{-\infty} \dot{\gamma} dt$.
For a propagating front with invariant shape, we have $\dot{\gamma} = \frac{\partial v}{\partial x} =  \frac{\partial v}{U_\text{f} \partial t}$, so  that
\begin{equation}
\gamma_\infty =  \frac{1}{U_\text{f}} \int^{U_0}_{0} dv =  \frac{U_0}{U_\text{f}} = \frac{1}{k}.
\label{eq:gamma_k}
\end{equation}
Thus, how fast a jamming front propagates depends on how much total strain is accumulated locally as the jamming front moves through. The physical picture is that a finite strain is required to shear the particles into an arrangement that jams. This suggests that the front propagation speed can depend on the system preparation conditions as well as on the straining history. A front will propagate faster (slower) if the suspension has been slowly pre-sheared along the same (opposite) direction, as confirmed in Supplementary Information.

\subsection*{Constitutive relation between applied stress and front speed}
 A central quantity of materials is constitutive relation connecting the applied stress $\Sigma$ to the velocity $U_0$. In our geometry, $\Sigma$ is readily extracted from momentum conservation. The momentum of an elongating jammed part of the material of cross sectional area $S$ is $p=\rho S x_f U_0$. Equating the time derivative of this quantity with the force $\Sigma S$, one obtains:
\begin{equation}
\Sigma = \rho U_0 U_\text{f} =   \rho k U_0^2,
\label{eq:Stress}
\end{equation}
where $\rho = 1.63 \times 10^3$~kg/m$^3$ is the density of the suspension. We calculated $\Sigma$ with Eq.~\ref{eq:Stress} using the experimental data. As Fig.~\ref{Compare}C shows, the dependence of $\Sigma$ on $U_0$ matches a quadratic power law well. Note that this stress level is above the upper limit of the stress in steady-state rheology experiments, which is provided by the surface tension (about 500~Pa) \cite{DST_review}. This suggests that the dilation of the material and the associated breaking of the air-liquid interface expected beyond this stress level does not have enough time to take place during the transient front propagation. This effect may however be apparent in the departure of the power-law behavior in Fig.~\ref{Compare}C for the largest stresses we probed, or equivalently (according to Eq.\ref{eq:Stress}) to the erosion of the plateau behavior of $k$ in Fig.~\ref{Compare}A. Indeed the theory developed below, which neglects this dilation effect, predicts a constant  behavior for $k(U_0)$.

Interestingly, the form of the constitutive relation Eq.\ref{eq:Stress} is identical to the expression for the dynamic pressure in a normal fluid, except that the density is renormalized by a factor $k$, so the effective density becomes $\rho_\text{eff} \sim k \rho$. Since $k$ can be as big as a factor of 10 according to our experiments, running on cornstarch is like running on a liquid ten times denser than water, thus generating a ten times larger lift force for the same motion of the legs. This ``added mass'' generated by the propagating fronts was tracked and imaged in previous impact experiments \cite{Scott, Ivo_2D, Endao}.

\subsection*{Maximum shear rate} The maximum shear rate $\dot{\gamma}_\text{max}$ characterizes the steepness of the fronts.
It is inversely related to the front width $\Delta$ since dimensionally we must have $ \Delta \sim U_0 / \dot{\gamma}_\text{max}$.
Fig.~\ref{Compare}D shows $\dot{\gamma}_\text{max}$ as a function of $U_0$. The experimental data for different $\phi$ from 0.462 to 0.532 collapse to very good approximation onto a single curve, revealing a power law of the form 
$ \dot{\gamma}_\text{max} \propto U_0^b $, with exponent $b_\text{exp} = 1.51 \pm 0.09$.  From this observation we can deduce that $ \Delta \propto U_0^{1-b} \propto U_0^{-0.5}$. Since $\dot{\gamma}_\text{max}$ is roughly independent of $\phi$, we predict the front width to be insensitive to $\phi$, which is different from what was found in the compression front \cite{Scott_EPL}.

\subsection*{A model for transient phenomena}

\begin{figure*}[!t]
\begin{center}
\includegraphics[scale = 1]{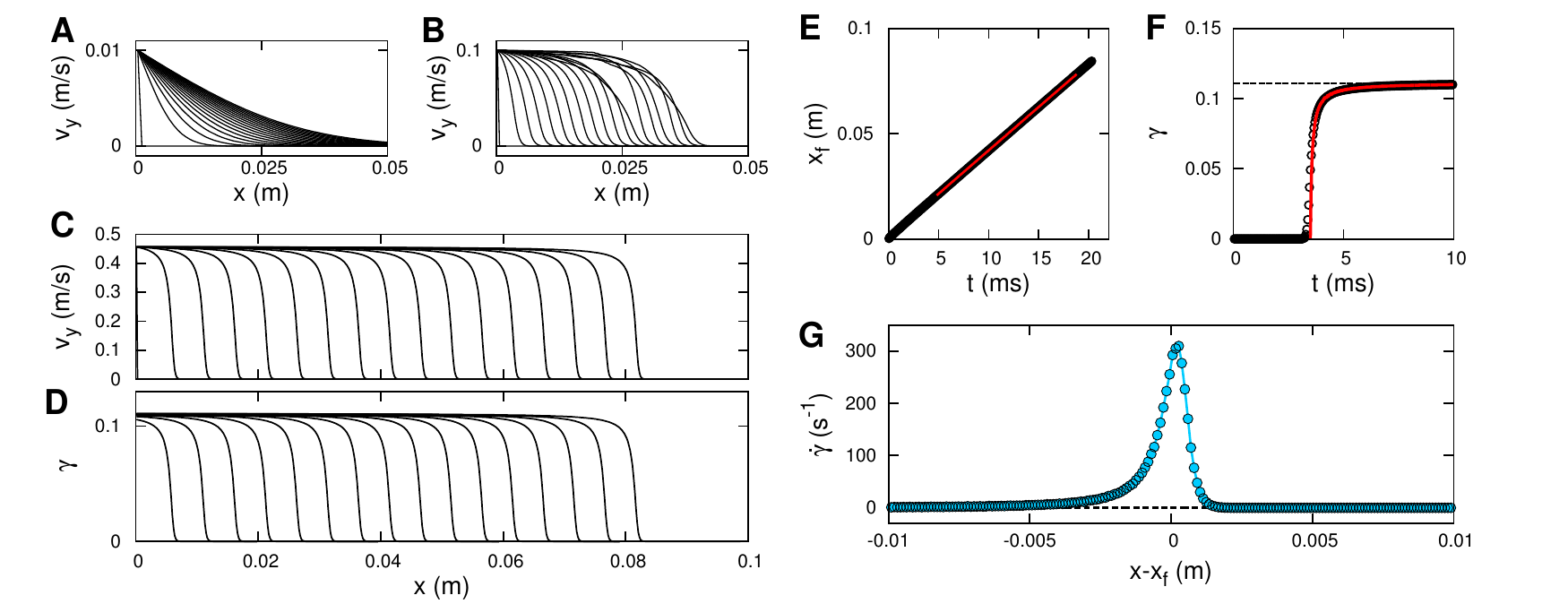}
\end{center}
\caption{1D model system for numerical calculation. For all data shown, the parameters are: $\phi_0 = 0.593$, $\phi_\text{m} = 0.452$, $\eta_0 = 13.6$~mPa$\cdot$s, $\Sigma^* = 20.4$~Pa, and $\gamma^* = 0.197$. (A, B) Velocity profiles at different times for $\phi = 0.521$, $U_0 = 0.01$~m/s (A) and 0.1~m/s (B). (A) is in the fluid-like regime, (B) is in the unstable regime. (C - G) Results in the jamming regime, for $\phi = 0.532$ and $U_0 = 0.456$~m/s. The results can be directly compared with the experiment shown in Fig.~\ref{Front_EXPT}. (C, D) Velocity profiles and accumulated strain $\gamma$ at different times. (E) Front position $x_\text{f}$ as a function of time. The red line shows a linear fit. (F) Accumulated strain $\gamma$ as a function of time in element $n = 80$. The red curve is a fit to a power law. The dashed black line indicates the asymptotic strain. (G) Local shear rate calculated from the mean velocity profile. }
\label{Front_CAL}
\end{figure*}

The Wyart-Cates model \cite{Wyart_Cates} describes shear thickening of suspensions under steady-state shear. The central ideas are that  (i)  if particles have a short range repulsion (due to charges,  polymer brushes, etc...), frictional contacts between them will be made only beyond a characteristic particle pressure $P^*$. The fraction of frictional  contacts $f(P)$ must be a growing function of $P$, such as:
\begin{equation}
\label{f}
f(P) = 1-\mathrm{exp}(-P/P^*).
\end{equation}
(ii) The packing fraction $\phi_\text{eff}$ at which jamming occurs is known to depend on the friction coefficient. For such particles it must then depend on $P$, as can be captured in a linear interpolation:
\begin{equation}
\phi_\text{eff}(P) = f(P) \phi_\text{m} + \left[ 1-f(P) \right] \phi_0,
\label{eq:phi_eff_steady}
\end{equation}
where $\phi_0$ and $\phi_\text{m}$ are the frictionless and frictional jamming packing fractions, respectively. 
(iii) When a suspension with packing fraction $\phi$ is under shear, the ratio between normal stress $P$ and shear rate $\dot{\gamma}$ diverges at $\phi_\text{eff}$: 
\begin{equation}
\label{div}
P / \dot{\gamma} \propto \left[ 1 - \phi / \phi_\text{eff}(P) \right]^{-\alpha}.
\end{equation}
For frictionless particles the exponent $\alpha$ can be computed analytically, leading to $\alpha = 2.85$ \cite{DeGiuli15a}, whereas for frictional particles it is smaller ~\cite{Trulsson16}. Here we pick $\alpha = 2$ for simplicity, a value also in good agreement with previous experimental results \cite{Reverse_Blanc, Boyer, Poon_Guy}. Recently, a similar but more detailed model has been validated with numerical simulations by Singh {\sl et al.} \cite{Morris2017}.

Eqs.~\ref{f},\ref{eq:phi_eff_steady},\ref{div} allow one to compute $P(\dot\gamma)$, eventually leading to a phase diagram predicting CST, DST and jamming in the $(\phi,\dot\gamma)$ plane. A further prediction of the shear stress can be obtained following the relation $ \Sigma = \mu P$, where $\mu$ is the macroscopic friction. It can in principle depend on $P$ and $\phi$, but in practice the dependence is mild on $\phi$ and essentially inexistent on $P$, and $\mu$ can thus be approximated as a constant.  

However, this model only applies to steady state. To model transient phenomena, consider an initial isotropic state where particles are not touching. There must be a characteristic strain $\gamma^*$ beyond which the microscopic structure becomes anisotropic and particles start to make contact. Let us denote the fraction of such particles by $g(\gamma)$, whose  contacts can be frictional or not (if the force is insufficient). $g(\gamma)$ must be a growing function, such as: 
\begin{equation}
g(\gamma) = 1-\mathrm{exp}\left( - \gamma / \gamma^* \right). 
\label{eq:g}
\end{equation}
The density of frictional contacts can now be estimated as $g(\gamma) f(\Sigma)$. We thus obtain for the jamming packing fraction: 
\begin{equation}
\phi_\text{eff}(\Sigma, \gamma) = g(\gamma) f(\Sigma) \phi_\text{m} + \left[ 1-f(\Sigma)g(\gamma) \right] \phi_0,
\label{eq:phi_eff}
\end{equation}
where $f(\Sigma) = 1-\mathrm{exp}(-\Sigma / \Sigma^*)$ and $\Sigma^*=\mu P^*$.   Making the additional approximation\footnote[2]{This is clearly a simplification, as the viscosity should not only depend on the fraction of frictional contacts, but also  on the anisotropy of the contact network characterized by $\gamma$.  Our results support that this dependence is not essential to describe fronts.} that in the transient as well, the viscosity only depends on  $ \phi_\text{eff}$   we obtain for the shear stress (in the spirit of  Eq.\ref{div}): 
\begin{equation}
\Sigma = \eta_0 \dot{\gamma} \left[ 1 - \phi / \phi_\text{eff}(\Sigma, \gamma) \right]^{-2},
\label{eq:S}
\end{equation}
where $\eta_0$ is the solvent viscosity. Note that this equation can be applied to higher dimensions as well, where
$\Sigma $ and $\dot{\gamma}$  now indicate the shear stress  and shear strain tensors respectively \footnote[3]{To describe propagating fronts in two or three dimensions, one may further assume that $\phi$ is constant in space since particle migration is slow, and use that the material is incompressible.}.

Eqs.~\ref{eq:g},\ref{eq:phi_eff},\ref{eq:S} lead to a closed relationship for $\Sigma(\gamma,\dot\gamma,\phi)$. 
If the suspension does not jam, we can take the limit $\gamma \to +\infty$ and Eq.~\ref{eq:phi_eff} reverts back to Eq.~\ref{eq:phi_eff_steady} for a steady-state system, as it should. For spatially non-uniform situations as fronts, 
Newton's law must be included. In the one-dimensional situation studied here this written as: 
\begin{equation}
\rho \frac{\partial v_y(x,t)}{\partial t} = \frac{\partial \Sigma}{\partial x}. 
\label{eq:EOM}
\end{equation}

\subsection*{Qualitative predictions of the model}
When $U_0$ (or equivalently the stress $\Sigma$) is sufficiently small, $\phi_\text{eff} \approx \phi_0$ and the viscosity is constant according to Eq.\ref{eq:S}. Injecting this relation into Eq.~\ref{eq:EOM} leads to a diffusion equation, and one recovers the usual flow profile for a liquid, evolving with a characteristic length scale $\sqrt{\nu t}$ toward a steady-state shear flow, where $\nu$ is the kinematic viscosity. We do recover such a diffusive profile in our finite element implementation of the model (see Supplementary Informations for numerical details), as shown in Fig.~\ref{Front_CAL}A.

By contrast, if $U_0$ is large and $\phi>\phi_m$ then there must exist a front separating a jammed and a flowing region. According to Eq.\ref{eq:Stress}, this front must move at constant speed. This is also recovered in our numerics. Fig.~\ref{Front_CAL} shows the numerically obtained velocity (C) and local accumulated strain (D) profiles at different times, as well as the front location $x_\text{f}$ (E),  the accumulated strain $\gamma$ at a fixed position(F), and local shear rate $\dot{\gamma}$ (G). The parameters are indicated in the caption, and they were chosen (see below) to correspond to the experimental data in Fig.~\ref{Front_EXPT}, allowing a direct  comparison. 
The model reproduces a front that propagates with a constant speed (C and E). The local accumulated strain always approaches a finite value asymptotically (D and F), which is in close agreement with observations. 
The shape of the $\dot{\gamma}(x)$ curve plotted in panel G also agrees with the experimental data in several key aspects. The maximum shear rate is obtained near $x = x_\text{f}$, and both curves show asymmetry with respect to $|x|-|x_\text{f}| = 0$: $\dot{\gamma}$ grows quickly as the front approaches, but decays with a tail after the front has passed by. However, the front is sharper than in the experiments, as quantified below.

Finally, at intermediate $U_0$ the model predicts a regime of instability (not seen in our experiments), exemplified in Fig.~\ref{Front_CAL}B. After propagating across a certain distance, the shape of the velocity profile in the co-moving frame oscillates back and forth. 
Such instability is not entirely surprising: for these velocities, the stress $\Sigma$ lies inside the S-shaped portion of the flow curves 
(see Supplementary Information). In that stress range a complex sequence of instabilities and chaotic behavior in steady-state systems has been reported experimentally \cite{Poon_Hermes}, which appears to be sensitive to the presence of a free surface that can be deformed. Modeling  the front in this velocity regime in the one-dimensional geometry discussed here may thus require to allow for deformation of the free surface. This goes beyond this work, and here we focus on the large $U_0$ regime.  

 \subsection*{Quantitative comparison with experiments}

There are five parameters  in our model, but we can obtain $\phi_0$, $\phi_\text{m}$, $\eta_0$, and $\Sigma^*$ from steady-state rheology. This is shown in the Supplementary Information where we obtain $\phi_0 = 0.593$, $\phi_\text{m} = 0.452$, $\eta_0 = 13.6$~mPa$\cdot$s, and $\Sigma^* = 20.4$~Pa. We are left with a single parameter, $\gamma^* = 0.197 \pm 0.002$, obtained by fitting the front propagation speed $k$ and its inverse $\gamma_\infty$ at different $\phi$, as shown in Fig.~\ref{K_SS_phi}- a very respectable agreement for a single parameter fit. Interestingly, $\gamma^* \approx 0.2$ is also found in regular granular materials \cite{Pailha} and in suspensions \cite{Sayantan} as the strain scale below which transient, start-up behavior is observed.

Note that the most important predicted quantities  ($k$ and  $\gamma_\infty$) can be estimated analytically in our model in the limit of large $U_0$. In that case, the stress is large when the front passes and we may take $f(\Sigma)\approx 1$ in Eq.\ref{eq:phi_eff}. Jamming occurs when $\phi_\text{eff}=\phi$, leading to $g(\gamma^*)=(\phi_0-\phi)/(\phi_0-\phi_\text{m})$. For our choice of $g$ this implies: 
\begin{equation}
\gamma_\infty = \gamma^* \cdot \text{ln} \frac{\phi_0 - \phi_\text{m}}{\phi - \phi_\text{m}}.
\label{eq:FP}
\end{equation}

To further test the model we compute $k$, $\gamma_\infty$, $\Sigma$ and $\dot{\gamma}_\text{max}$ across a range of packing fractions $\phi$ and boundary speeds $U_0$, and compare the results with experiments directly in Fig.~\ref{Compare}. 
As follows from Eq.\ref{eq:FP}, we predict  $k$ and $\gamma_\infty$ to be essentially independent of $U_0$ for large values, and 
$\Sigma\sim U_0^2$. These predictions match the data very well at each $\phi$ (except for the largest $\phi$ values where $k$ shows some decay, presumably induced by the deformation of the free interface as discussed above).

As shown in Fig.~\ref{Compare}D, $\dot{\gamma}_\text{max}$ obtained from experiments (solid circles) and calculations (hollowed circles) both obey power laws as functions of $U_0$, and their pre-factors are both relatively $\phi$-independent over the range $\phi \in [0.462, 0.532]$ (see Supplementary Information). However, the model predicts an exponent around 2 instead of 1.5, and the pre-factor is about one order of magnitude larger. More sophisticated models describing not only the fraction of frictional contacts, but also the evolution of the anisotropy of the contact network with strain, may be required for a detailed treatment of the front width.

\subsection*{Conclusions}

We showed experimentally that when a dense suspension in its quiescent, unjammed state is suddenly sheared by moving one of its boundaries, a rapidly propagating jamming front can be initiated that transforms the suspension from a fluid-like state into a solid-like state. We found that the properties of such fronts are controlled by the locally accumulated shear strain. These transient, start-up dynamics can be captured by introducing a characteristic strain scale $\gamma^*$ into the Wyart-Cates model originally developed to describe the steady-state rheology of shear-thickening suspensions. Despite its simplicity, this extended model gives very good agreement with the experiments, quantitatively reproducing the dependence of the normalized fronts speed  $k$ and of the locally accumulated shear strain $\gamma_\infty$ on  packing fraction $\phi$. It also predicts correctly the qualitative dependence  on system and forcing parameters of the maximum shear rate $\dot{\gamma}_\text{max}$ inside the front. 

Importantly, the generalized Wyart-Cates model introduced here establishes a direct link between the steady-state and transient behaviors in dense suspensions. It shows that to obtain jamming fronts, the packing fraction of the suspension must be above the frictional jamming packing fraction $\phi_\text{m}$. In the range between $\phi_\text{m}$ and the frictionless jamming packing fraction $\phi_0$, the suspension will evolve into a state that jams at high stress, but can still flow at low stress. 

While we discussed the model in its simplest form, appropriate for a semi-infinite 1D system, the same ideas and numerical approaches should allow for several extensions. This includes accounting for the presence of walls (which can take up large stresses once reached by the fronts) as well as extension to 2D or 3D systems (where the fronts propagate with different speeds in the directions along the applied forcing and perpendicular to it \cite{Ivo_2D, Endao, Sayantan}).


\subsection*{Materials and Methods}
In the rheology experiments we used suspensions of cornstarch (Ingredion). The dry cornstarch particles were stored in a temperature and humidity controlled environment at $22.5 \pm 0.5^\circ$C and $44 \pm 2 \%$ relative humidity. The solvent was a mixture of caesium chloride (CsCl), glycerol and deionized water. The mass ratio between glycerol and water in the solvent was $65\%:35\%$. The density of the solvent was $1.62 \times 10^3$~kg/m$^3$, which matched the density of cornstarch particles to prevent sedimentation. The viscosity of the solvent was $11 \pm 1$~mPas. When a suspension was made, we mixed $m_\text{cs}$ grams of cornstarch particles with $m_\text{l}$ grams of the solvent and left it sit still for approximately 2 hours before performing experiments to allow full wetting of the particles and for most air bubbles to disappear. The packing fraction $\phi$ of the suspension was calculated by
\begin{equation}
    \phi = \frac{1}{1-\psi} \frac{(1-\xi) m_\text{cs}/\rho_\text{cs}}{(1-\xi) m_\text{cs}/\rho_\text{cs}+m_\text{l}/\rho_\text{l}+\xi m_\text{cs}/\rho_\text{w}},
    \label{eq:phi_m_cs}
\end{equation}
where $\rho_\text{cs}$ and $\rho_\text{l}$ represent the density of the particles and the solvent, respectively, $\rho_\text{w}$ is the density of water, $\xi$ is the mass ratio of moisture in the cornstarch particles in our lab environment, and $\psi$ is the porosity of cornstarch particles. We used $\xi = 0.13$, $\psi = 0.31$, and $\rho_\text{cs} = 1.63 \times 10^3$~kg/m$^3$ in the calculation of $\phi$ \cite{Endao_SOS}.

\subsection*{Acknowledgements}
We thank Tonia Hsieh for providing the linear actuator. We thank Eric Brown, Mike Cates, Yoel Forterre, Nicole James, Bloen Metzger, Kieran Murphy, Christopher Ness, Olivier Pouliquen, John Royer, and Adam Wang for many useful discussions. This work was supported by the US Army Research Office through grant W911NF-16-1-0078, the Swiss National Science Foundation under Grant No. 200021-165509 and the Simons Foundation Grant ($\#$454953 Matthieu Wyart). IRP acknowledges financial support from the Royal Society through grant RG160089. Additional support was provided by the Chicago MRSEC, which is funded by NSF through grant DMR-1420709.


\clearpage
\newpage

\renewcommand{\theequation}{S\arabic{equation}}
\renewcommand{\figurename}{Fig.~S}

\setcounter{equation}{0}
\setcounter{figure}{0}

{\bf{\Large Supplementary Information}}


\subsection*{Flow profiles for slow $U_0$}

\begin{figure*}
\begin{center}
\includegraphics[scale = 1.3]{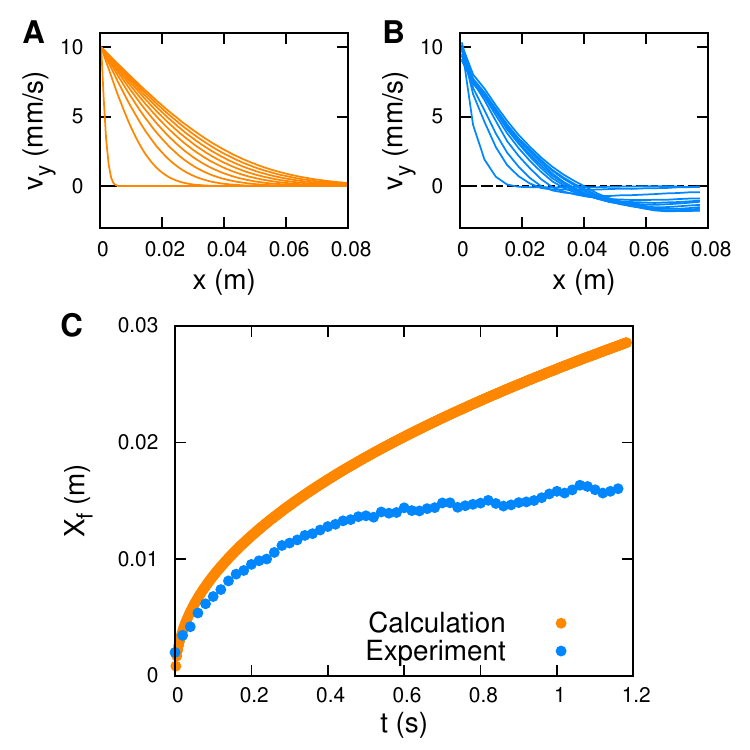}
\end{center}
\caption{\label{SlowProfile} Flow profiles in the fluid-like regime and ``front'' position $x_\text{f}$ at $\phi = 0.521$ and $U_0 = 0.01$~m/s. (A) Numerical calculation based on the model. (B) Experimental data. (C) Position of the front, defined as the position where $u_\text{y} = 0.45 U_0$. 
}
\end{figure*}

When $U_0$ is sufficiently slow, the suspension is in the lubrication regime and behaves like a Newtonian fluid. For a Newtonian fluid sheared in a semi-infinite 1D system, the flow profile is self-similar with a characteristic length scale $\sqrt{\nu t}$, where $\nu$ is the kinematic viscosity. If we define a normalized, time-dependent length scale $s = x/(\sqrt{\nu t})$, the velocity $u(x,t)$ is \cite{SI_Fluid}: 
\begin{equation}
u(x,t) = U_0 \left[ 1-\mathrm{erf}(s/2) \right], 
\label{eq:NF}
\end{equation} 
where $\text{erf}(x)$ is the error function. The numerically calculated and experimentally measured flow profiles at $\phi = 0.521$ and at sufficiently slow speed $U_0 = 0.01$~m/s are shown in Fig.~S\ref{SlowProfile} as an example. One major difference is that in the calculation the system is strictly one-dimensional, so the local velocity is always positive during the whole process. However, in the experiment negative flow velocity is observed further away from the plate, which originates from fluid re-circulation due to the finite container size. 
We can still define the ``front position'' $x_\text{f}$ as the $x$ position at which $u = 0.45 U_0$. As shown in Fig.~S\ref{SlowProfile}C, in the calculation $x_\text{f}$ keeps growing as a function of time, and before the flow reaches the other boundary it satisfies  
$ x_\text{f} \propto \sqrt{\nu t} $,  
where $\nu = \eta_0 (1-\phi/\phi_0)^{-2} / \rho$. In the experiments, the front almost stopped propagating at late time and reached a steady state. As a result, in experiments with slow $U_0$, we obtain $U_\text{f} \approx 0$. \\

\subsection*{Effect of pre-shear}

\begin{figure*}
\begin{center}
\includegraphics[scale =  1.3]{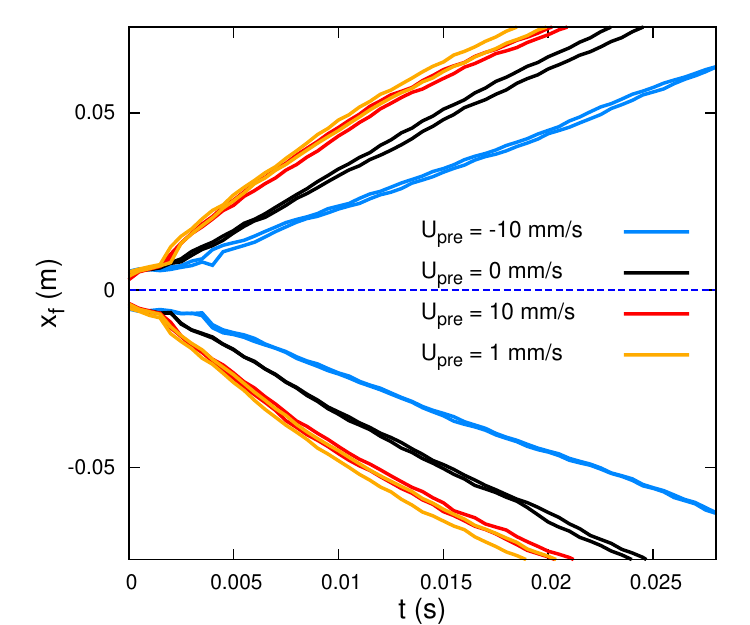}
\end{center}
\caption{\label{k_preshear} Front position as a function of time for different pre-shear. The fast plate speed was $U_0 = 360$~mm/s, and the slow pre-shear speed $U_\text{pre}$ varied as labeled in the plot. Positive $U_\text{pre}$ represents pre-shear in the same direction as $U_0$, and negative $U_\text{pre}$ was in the opposite direction. }
\end{figure*}

To prepare a system with non-zero initial strain, we applied a pre-shear at a slow speed $U_\text{pre}$, where the suspension is still fluid-like. For testing the effect of pre-shear we moved the plate 10~mm forward or backward at $U_\text{pre} = 1$~mm/s or 10~mm/s, and then applied fast shear at $U_0$. Results for $\phi = 0.526$ and $U_0 = 0.36$~m/s are shown in Fig.~S\ref{k_preshear} as an example. We performed pre-shear at different $U_\text{pre}$ (from 0.1~mm/s to 10~mm/s), and waited for different lengths of time between pre-shear and fast shear, from several seconds to 10 minutes. In each case we obtained almost identical $x_\text{f}$-$t$ curves, as long as $U_\text{pre}$ was slow enough so that the suspension remained fluid-like. This also shows that cornstarch suspensions can be treated as an athermal system over time scales as long as several minutes. 

Note that in our experiments the velocity profile was not always linear during the pre-shear. This was due to the limited range the plate could move, so the distribution of ``pre-strain'' was not exactly the same everywhere. When the pre-shear finished, the accumulated strain close to the plate was the maximum and it decreased gradually to the side. As a result, in the following step, when pushed with a fast speed $U_0$, the front speed $U_\text{f}$ slowed down as it propagates away from the plate, which can be seen in Fig.~S\ref{k_preshear}. 
All these observations support our argument that the front propagation speed is dependent on the initial configuration and arrangement of the particles. \\

\subsection*{Steady-state rheology experiments} 

\begin{figure*}
\begin{center}
\includegraphics[scale=0.8]{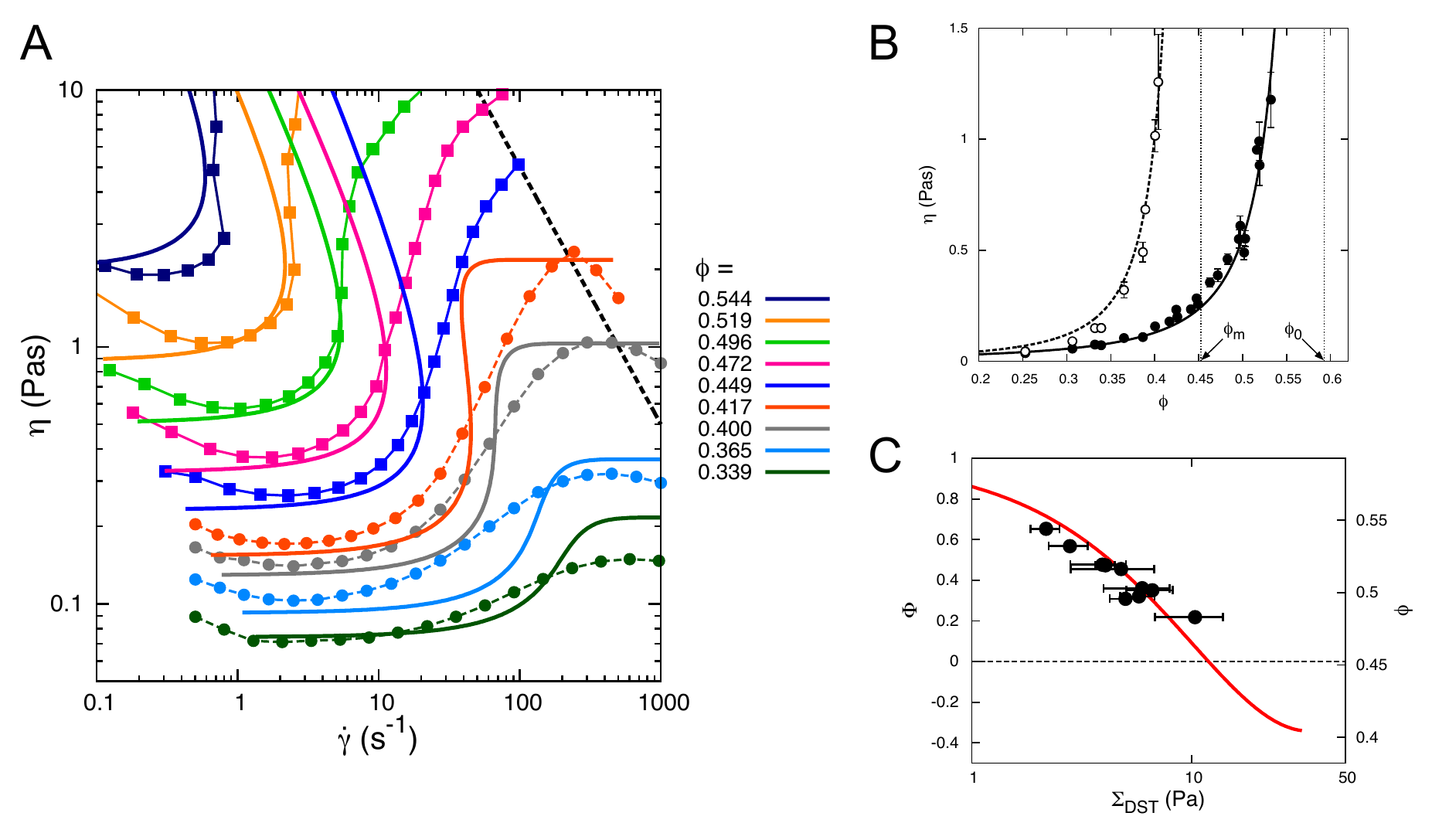}
\end{center}
\caption{\label{Rheology} 
(A) Viscosity $\eta$ at different shear rates $\dot{\gamma}$ and packing fractions $\phi$. Squares connected by thin solid lines represent stress controlled experiments; circles connected by thin dashed lines represent shear rate controlled experiments. The predictions of the Wyart-Cates model are shown by the thick curves with the same color as the experiments. The dashed black line indicates a constant stress $\Sigma = 500$~Pa, which is provided by surface tension and corresponds to the upper limit of stress in steady-state experiments using our shear cell geometry. 
(B) The lower Newtonian viscosity $\eta_{\text{N},1}$ (solid circles) and higher Newtonian viscosity $\eta_{\text{N},2}$ (open circles) at different $\phi$. The two curves show the best fit of $\eta_{\text{N},1}$ and $\eta_{\text{N},2}$ with Eq.~\ref{eq:ViscN_newFit}. The vertical dashed lines label $\phi_\text{m}$ (left) and $\phi_0$ (right) obtained from the fitting. 
(C) Relation between rescaled packing fraction $\Phi$ and onset stress $\Sigma_\text{DST}$. For each $\Phi$, the corresponding $\phi$ is labeled on the right. The solid black points are experimental data. The red curve is the prediction of the model for $\Sigma^* = 20.4~$Pa. }
\end{figure*}

The steady state rheology experiments were performed with an Anton Paar MCR 301 rheometer. The suspensions were tested between parallel plates, and the diameter of the upper plate was 25~mm (tool PP25). An enclosed solvent trap was used to prevent evaporation. We performed both shear rate controlled and shear stress controlled experiments at different $\phi$. Before each measurement, the suspension was pre-sheared by ramping from $\Sigma = 0.1$~Pa to 100~Pa for 50 s in total, then sheared slowly at $\Sigma = 0.1$~Pa for 30~s to 60~s. After these two steps of preparation, we ran the actual measurements, where we took 20 data points in a scan from low to high $\dot{\gamma}$ or $\Sigma$ (from approximately 0.1~Pa to 1000~Pa). At each point the measurement lasted for 10~s to 30~s, and we made sure that the time was long enough so that the viscosity did not vary with time. Some exemplary viscosity-shear rate data ($\eta$-$\dot{\gamma}$ curves) at different $\phi$ are shown in Fig.~S\ref{Rheology}A.


The Wyart-Cates model predicts that, for suspensions in the CST and DST regimes, the $\eta$-$\dot{\gamma}$ curves have two Newtonian plateau: $\eta_{\text{N},1}$ at low stress and $\eta_{\text{N},2}$ at high stress. Both $\eta_{\text{N},1}$ and $\eta_{\text{N},2}$ increase with $\phi$, and the stress threshold $\Sigma^*$ controls the stress at which the transition occurs from one plateau to the other. In the experiments there are several differences from this model, which we need to account for. 
Firstly, dense suspensions show shear thinning at low shear rate. To accommodate this we took the average viscosity in the flat section near the minimum of a $\eta$-$\dot{\gamma}$ curve as $\eta_{\text{N},1}$. 
Secondly, the higher branches of the $\eta$-$\dot{\gamma}$ curves are more like smooth peaks instead of plateaus. We therefore took the peak values of $\eta$ as $\eta_{\text{N},2}$. 
Lastly, in steady-state rheology experiments there is another stress limit set by the surface tension at the suspension-air interface, which confines the suspensions between the parallel plates. The empirical relation is $\Sigma_\text{max} \approx 0.1 \Gamma/d$, where $\Gamma$ is the surface tension of the solvent and $d$ is the particle diameter \cite{SI_DST_review}. The surface tension of our solvent was about 75~N/m \cite{SI_SurfaceTension_1, SI_SurfaceTension_2} and the average diameter of cornstarch granules was about 15~$\mu$m \cite{SI_CS_Size}. As a result, $\Sigma_\text{max}$ was 500~Pa approximately. Above this stress the surface tension could not confine the suspension and the measurements became unreliable, \textit{i.e.} the data could no longer be used to extract $\eta_{\text{N},2}$. 

According to the Wyart-Cates model, the viscosity of a suspension is
\begin{equation}
\eta = \frac{\Sigma}{\dot{\gamma}} = \eta_0 \left[ 1-\phi/\phi_\text{eff}(\Sigma) \right]^{-2}. 
\label{eq:eta}
\end{equation}
In the two limits of $\Sigma$, Eq.~\ref{eq:eta} has two asymptotes  
\begin{equation}
\begin{split}
\eta_{\text{N},1} &= \eta_0 (1-\phi/\phi_0)^{-2},~(\Sigma \to 0), \\ 
\eta_{\text{N},2} &= \eta_0 (1-\phi/\phi_\text{m})^{-2},~(\Sigma \to +\infty). 
\end{split}
\label{eq:ViscN_newFit}
\end{equation} 
This predicts that though both $\eta_{\text{N},1}$ and $\eta_{\text{N},2}$ increase with $\phi$, they grow with different rate and diverge at different $\phi$: $\eta_{\text{N},2}$ diverges at $\phi = \phi_\text{m}$ while $\eta_{\text{N},1}$ diverges at $\phi_0$. 
Fig.~S\ref{Rheology}B shows $\eta_{\text{N},1}$ and $\eta_{\text{N},2}$ as functions of $\phi$. We fit both $\eta_{\text{N},1}$ and $\eta_{\text{N},2}$ simultaneously on log scales to Eq.~\ref{eq:ViscN_newFit} and obtain the parameters $\eta_0 = 13.6$~mPas, $\phi_0 = 0.593$ and $\phi_\text{m} = 0.452$.


We can then use the onset stress of DST, $\Sigma_\text{DST}$, to obtain the threshold stress $\Sigma^*$. $\Sigma_\text{DST}$ is the stress at the turning point where a $\eta$-$\dot{\gamma}$ curve becomes vertical, so we have 
\begin{equation}
\frac{d \dot{\gamma}}{d \Sigma}\bigg\rvert_{\Sigma = \Sigma_\text{DST}} = 0.
\label{eq:slope}
\end{equation}  
Now with the three parameters $\eta_0$, $\phi_0$ and $\phi_\text{m}$ already extracted, $\Sigma_\text{DST}$ is only a function of $\phi$ and $\Sigma^*$. Equivalently, we can use a rescaled packing fraction $\Phi$, defined as 
\begin{equation}
\Phi = \frac{\phi - \phi_\text{m}}{\phi_0-\phi_\text{m}}. 
\label{eq:Phi}
\end{equation} 
Fig.~S\ref{Rheology}C shows the relation between $\Phi$ and $\Sigma_\text{DST}$ obtained from experiments. 
To obtain the $\Sigma^*$ that best fits $\Phi$-$\Sigma_\text{DST}$, we varied $\Sigma^*$ from 15~Pa to 25~Pa. For each $\Sigma^*$, we calculated the $\Sigma$-$\dot{\gamma}$ curve and found the corresponding $\Sigma_\text{DST}$ at the experimentally measured packing fractions. Then we calculated the sum of squared residuals (SSR) between the measured and calculated $\Sigma_\text{DST}$, and obtained $\Sigma^* = 20.4$~Pa, for the minimum SSR.


The four parameters to describe the steady-state behavior of our suspensions are: $\eta_0 = 13.6 \times 10^{-3}$~Pas, $\phi_0 = 0.593$, $\phi_\text{m} = 0.452$, and $\Sigma^* = 20.4$~Pa. With these in hand, we can calculate the $\eta$-$\dot{\gamma}$ relation at any packing fraction with the Wyart-Cates model and compare it with the experimental measurement, as shown in Fig.~S\ref{Rheology}A. 
The lowest three curves (green, light blue and gray) are in the CST regime with $\phi < \phi_\text{m}$. The next two curves above, at $\phi = 0.417$ and 0.449, are in the DST regime where $\phi_\text{m} < \phi < \phi_0$. In this regime, one might expect to see a discontinuous jump in viscosity, while the transitions seen in the experiments are less sharp than the model predicts. We note that this ``sharpness'' may be affected by the size distribution of the particles. It has been shown that the onset stress of shear thickening is a function of the particle size \cite{SI_Poon_Guy}: The larger the particles, the smaller the onset stress. Since cornstarch is highly poly-disperse, there should be a distribution of onset stress in the system, which smooths the transition. Lastly, the curves at the four highest packing fractions (from 0.472 to 0.544) are in the jamming regime where $\phi > \phi_\text{m}$. \\

\subsection*{Front speed at fast $U_0$ and large $\phi$}

\begin{figure*}
\begin{center}
\includegraphics[scale = 1.6]{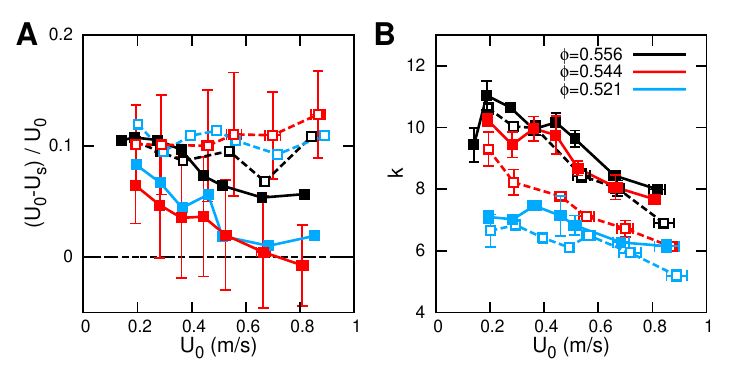}
\end{center}
\caption{\label{K_sandpaper} (A) Relative speed between the plate and suspension, $(U_0-U_\text{s}) / U_0$, at different $U_0$. Solid squares are from experiments with sandpaper, and open squares are without sandpaper. Blue: $\phi = 0.521$, red: $\phi = 0.544$, and black: $\phi = 0.556$. (B) Dimensionless front propagation speed $k$ at different $U_0$. The labels are the same as in (A). }
\end{figure*}

For most packing fractions, the normalized front speed $k$ is independent of shearing speed $U_0$ once $U_0$ becomes sufficiently large. However, for high packing fractions, $k$ is found to decrease at large $U_0$ (see Fig. 2 in the main text). We consider two possibilities: one is slip at the boundary, and the other is that the jammed suspension yields at high stress. When there is slip at the boundary, the actual speed of the suspension close to the boundary $U_\text{s}$ is slower than the driving speed $U_0$, so calculating $k$ using $U_0$ generates smaller predictions. To test this, we performed experiments with and without sandpaper (waterproof, grain size $\approx 10~\mu$m) at the plate-suspension interface and measured the relative speed difference $(U_0 - U_\text{s}) / U_0$. From the results shown in Fig.~S\ref{K_sandpaper}A, we can see that the rough surface did prevent boundary slip except for the highest packing fraction $\phi = 0.556$. Converting this to $k$ values as shown in Fig.~S\ref{K_sandpaper}B, experiments with boundary slip led to smaller $k$ in general, especially at $\phi = 0.544$. However, if we look at the experiments where $\phi = 0.544$, $U_0 > 0.4$~m/s and where $\phi = 0.521$, $U_0 > 0.5$~m/s, even though sandpaper minimized the boundary slip, the $k$ values still decreased significantly, and $\gamma_\infty$ increased correspondingly. This enhanced asymptotic strain likely is a sign of incipient yielding of the suspension. \\

\subsection*{Numerical calculations}

\begin{figure*}
\begin{center}
\includegraphics[scale=0.7]{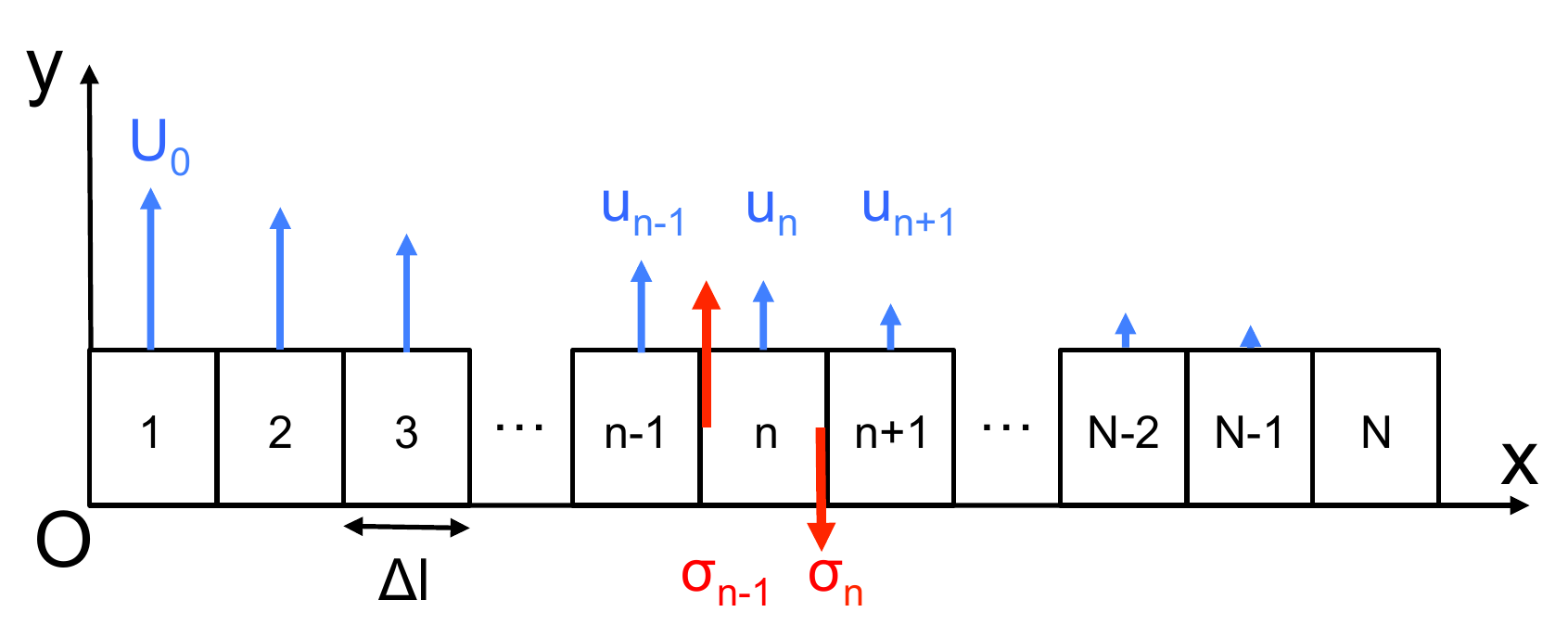}
\end{center}
\caption{\label{system} Schematic illustration of the model system used for the numerical calculations. The black boxes represent fluid elements, the blue arrows represent the local velocities and the red arrows show the shear stress applied on the left and right boundaries of the n-th element. The boundary conditions are $u_1 = U_0$ and $u_\text{N} = 0$. The width of an element is $\Delta l$. }
\end{figure*}

The one-dimensional model system we considered for the numerical calculations is illustrated in Fig.~S\ref{system}. It was comprised of $N$ elements aligned in the $x$ direction as labeled. Each element was allowed to move in the y direction only. The initial condition was zero velocity and zero strain for every element. At time $t = 0$~s the velocity of the 1st element was set to be $U_0$ and kept fixed throughout the calculation. The velocity of the $N$'th element was set to be zero for all times. For the other elements, the velocity was calculated using the forward Euler method. The equation of motion was: 
\begin{equation}
\Delta u_n =  \frac{\Delta t}{\rho \Delta l} (\sigma_{n-1} - \sigma_n), 
\label{eq:Du_num}
\end{equation} 
where $\rho$ is the density of the suspension, $u$ is the velocity of the element in the y direction, $\sigma$ is the shear stress applied on its boundaries, $\Delta l$ and $\Delta t$ are the length and time scales, respectively. 
From time step $i$ to $i+1$, we had 
\begin{equation}
u_n(i+1) = u_n(i) + \Delta u_{n}(i). 
\label{eq:u_new_num}
\end{equation} 
The stress $\sigma_n$ was calculated using 
\begin{equation}
\sigma_n = \eta_0 \dot{\gamma}_n [1 - \phi / \phi_{\text{eff},n}]^{-2}, 
\label{eq:P_num}
\end{equation} 
where 
\begin{equation}
\dot{\gamma}_n = \frac{u_{n} - u_{n+1}}{\Delta l}~(n \ge 2), 
\label{eq:SR}
\end{equation} 
and $\phi_{\text{eff}, n}$ was calculated according to the generalized Wyart-Cates model (Eq.~3, Eq.~6 and Eq.~7 in the main text of the paper). The increment of strain in every step was 
\begin{equation}
\gamma_n(i+1)  = \gamma_n(i) + \dot{\gamma}_n \Delta t. 
\label{gamma_num}
\end{equation} \\

\subsection*{Transition from slow to fast $U_0$}

\begin{figure*}
\begin{center}
\includegraphics[scale=1.3]{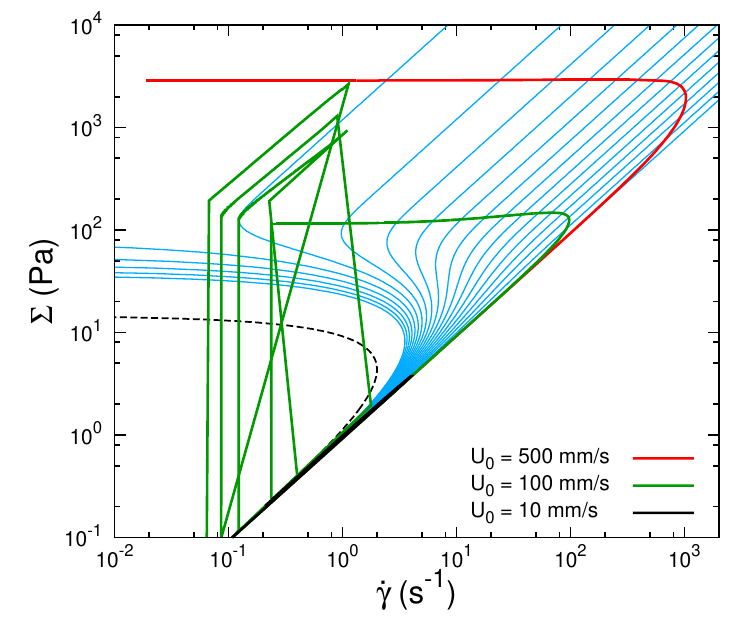}
\end{center}
\caption{\label{Instability} Evolution of $\Sigma$-$\dot{\gamma}$ at $\phi = 0.521$. The blue curves show the $\Sigma$-$\dot{\gamma}$ relation at different $\gamma$ (starting from zero, with strain increments of 0.0105 between adjacent curves), as predicted by the generalized Wyart-Cates model. The dashed black line corresponds to the relation at steady state ($\gamma \to +\infty$). The thick black, green and red lines show the relation between stress and shear rate in element no. 2, calculated numerically for different $U_0$ as indicated. }
\end{figure*}

To better understand the transition from slow $U_0$ to fast $U_0$, we look at the evolution of the $\Sigma$-$\dot{\gamma}$ relation as $\gamma$ accumulates, as shown in Fig.~S\ref{Instability}. The $\Sigma$-$\dot{\gamma}$ relation for a steady state system is labeled by the dashed black curve. Since $\phi > \phi_\text{m}$, it intersects with the $\dot{\gamma} = 0$~s$^{-1}$ axis, and does not have an upper branch. However, in the generalized model, since we introduced the g($\gamma$) term, the $\Sigma$-$\dot{\gamma}$ relation evolves as $\gamma$ accumulates. The $\Sigma$-$\dot{\gamma}$ relations at different $\gamma$ are presented by the blue curves in Fig.~S\ref{Instability}. When $\gamma = 0$, the relation between $\Sigma$ and $\dot{\gamma}$ is linear with a constant viscosity $\eta_0 (1 - \phi / \phi_0)^{-2}$. As $\gamma$ increases, the $\Sigma$-$\dot{\gamma}$ curve turns from linear to sigmoidal and finally approaches the black dashed line as $\gamma \to +\infty$. 

Given the $\Sigma$-$\dot{\gamma}$ relation at any $\gamma$, we now discuss, as a specific example, the variation of $\Sigma$ with $\dot{\gamma}$ in element no. 2 of the numerical 1D system, which we call the ``state'' of that element. When $U_0 = 0.01$~m/s, the state moves up along the Newtonian-fluid line and then turns back down along an almost identical path as $\dot{\gamma}$ is varied (black line). In contrast, at $U_0 = 0.5$~m/s the stress quickly reaches the upper branch of the sigmoidal curves and stays up there as $\gamma$ keeps accumulating and $\dot{\gamma}$ slows down. If $\gamma$ keeps increasing, the shear rate approaches $\dot{\gamma} = 0$~s$^{-1}$. Since $\Sigma$ stays constant, the viscosity of the suspension diverges as $\dot{\gamma} \to 0$. This then leads to a jammed state (red line).
At intermediate $U_0$ the system can enter a regime where the flows become unstable (green lines). Here the stress reaches the upper branch and forms a plateau at the beginning. However, as the strain accumulates and the strain rate slows down, the state of the element (at that stress level) enters a section of the S-shaped $\Sigma$-$\dot{\gamma}$ curves with negative slope. As a consequence, the stress has to jump down to the lower branch. The stress then builds up again and jumps back to the upper branch, and the process repeats.  \\

\subsection*{An alternative derivation of Eq.~10}

\begin{figure*}
\begin{center}
\includegraphics[scale=1.5]{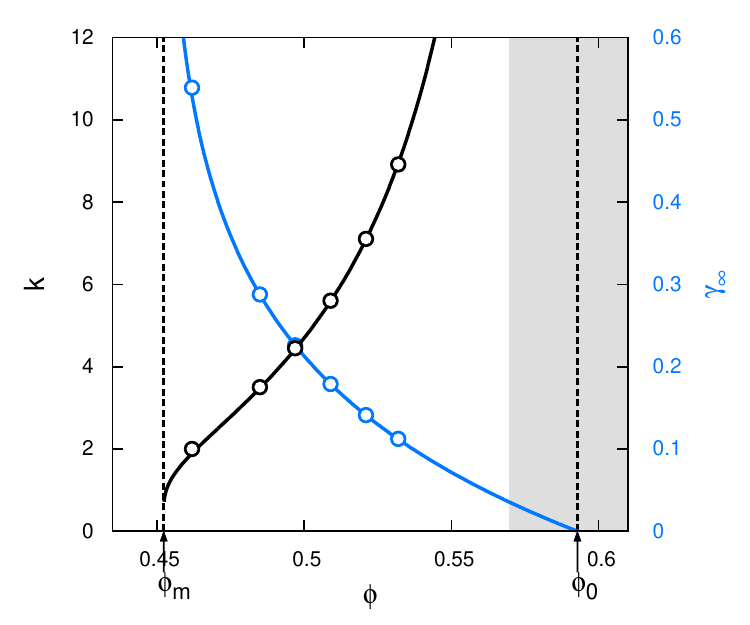}
\end{center}
\caption{\label{K_SS_CAL} Dimensionless front propagation speed $k$ and asymptotic accumulated strain $\gamma_\infty$ at different packing fraction $\phi$ obtained numerically at $\gamma^* = 0.197$ and $U_0 = 1~$m/s. The solid curves show Eq.~\ref{eq:FP} and its reciprocal at the same $\gamma^*$. }
\end{figure*}

Eq.~10 in the main text is an approximate relation between $\gamma_\infty$ and $\gamma^*$ in the regime of sufficiently fast $U_0$ where the front speed can be assumed constant. To keep the calculation simple, we make three approximations that are appropriate for this high speed limit: 
First, we approximate Eq.~8 in the main text by
\begin{equation}
\Sigma \approx \widetilde{\eta}_0 \cdot \dot{\gamma} \left[ \phi_\text{eff} - \phi  \right]^{-2},
\label{eq:sigma}
\end{equation} 
where $\widetilde{\eta}_0 \equiv \eta_0 \phi_0^2$. 
Second, in this limit $\Sigma$ is much larger than $\Sigma^*$, so we take $f(\Sigma) \approx 1$. 
Finally, since the front profile has an approximately invariant shape while propagating, the accumulated strain can be written as $\gamma(x,t) = \gamma (U_\text{f} t-x) \equiv \gamma(X)$. This leads to  
\begin{equation}
\gamma' \equiv \frac{d \gamma(X)}{d X} = \frac{1}{U_\text{f}} \frac{\partial \gamma}{\partial t} = -\frac{\partial \gamma}{\partial x}, 
\label{eq:F_p}
\end{equation} 
and 
\begin{equation}
\gamma'' \equiv \frac{d^2 \gamma(X)}{d X^2} = \frac{1}{U_\text{f}^2} \frac{\partial^2 \gamma}{\partial t^2} = \frac{\partial^2 \gamma}{\partial x^2}. 
\label{eq:F_pp}
\end{equation} 
Plugging Eq.~\ref{eq:sigma} into the equation of motion:
\begin{equation}
\rho \frac{\partial^2 \gamma}{\partial t^2} = \frac{\partial^2 \Sigma}{\partial x^2}, 
\label{eq:EOM_gamma}
\end{equation} 
we get
\begin{equation}
\rho \frac{\partial^2 {\gamma}}{\partial{t}^2} = \frac{\partial{^2}}{\partial{x}^2} \left\{ \frac{\widetilde{\eta}_0 \dot{\gamma}}{[(\phi_0 - \phi_\text{m}) e^{-\gamma/\gamma^*} + \phi_\text{m} - \phi]^2} \right \}. 
\label{eq:motion_2}
\end{equation} 
Using Eq.~\ref{eq:F_p} and Eq.~\ref{eq:F_pp}, we obtain 
$$ \frac{d^2}{d X^2} \left \{ \rho U_\text{f}^2 \gamma - \frac{\widetilde{\eta}_0 U_\text{f} \gamma'}{[(\phi_0 - \phi_\text{m}) e^{-\gamma / \gamma^*} + \phi_\text{m} - \phi]^2} \right \} = 0, $$ 
which leads to 
\begin{equation}
\rho U_\text{f}^2 \gamma - \frac{\widetilde{\eta}_0 U_\text{f} \gamma'}{[(\phi_0 - \phi_\text{m}) e^{-\gamma/\gamma^*} + \phi_\text{m} - \phi]^2} = C_1 X + C_2, 
\label{eq:motion_F}
\end{equation} 
where $C_1$ and $C_2$ are constants. In the region not yet reached by the front, both $\gamma$ and $\gamma'$ are zero. This means that as $X \to -\infty$ (at large $x$ or small $t$), the left hand side of Eq.~\ref{eq:motion_F} is zero, so the constants should be $C_1 = C_2 = 0$, and we obtain a first order equation governing the evolution of $\gamma$:
\begin{equation}
\frac{d\gamma}{dX} = \frac{\rho U_\text{f}}{\widetilde{\eta}_0} \gamma \cdot \left[ (\phi_0 - \phi_\text{m}) e^{-\gamma / \gamma^*} + \phi_\text{m} - \phi \right]^2.
\label{eq:DynamicEquation}
\end{equation}
It has two fixed points. For any given $x$, $\gamma$ increases with time from an unstable fixed point $\gamma = 0$ to a half-stable fixed point, which is the asymptotic accumulated strain:
\begin{equation}
\gamma_\infty = \gamma^* \cdot \text{ln} \frac{\phi_0 - \phi_\text{m}}{\phi - \phi_\text{m}}.
\label{eq:FP}
\end{equation}
Written as a function of the rescaled packing fraction $\Phi$ defined in Eq.~\ref{eq:Phi}, it becomes
\begin{equation}
\gamma_\infty = - \gamma^* \text{ln} \Phi.
\label{eq:FP_Phi}
\end{equation} 
This approximate result captures the relation between $\gamma_\infty$ and $\gamma^*$ very well. In Fig.~S\ref{K_SS_CAL} we compare the numerically calculated $k$ and $\gamma_\infty$ at $\gamma^* = 0.197$ and $U_0 = 1$~m/s with Eq.~\ref{eq:FP}. \\

\subsection*{Maximum shear rate}

\begin{figure*}
\begin{center}
\includegraphics[scale=1.5]{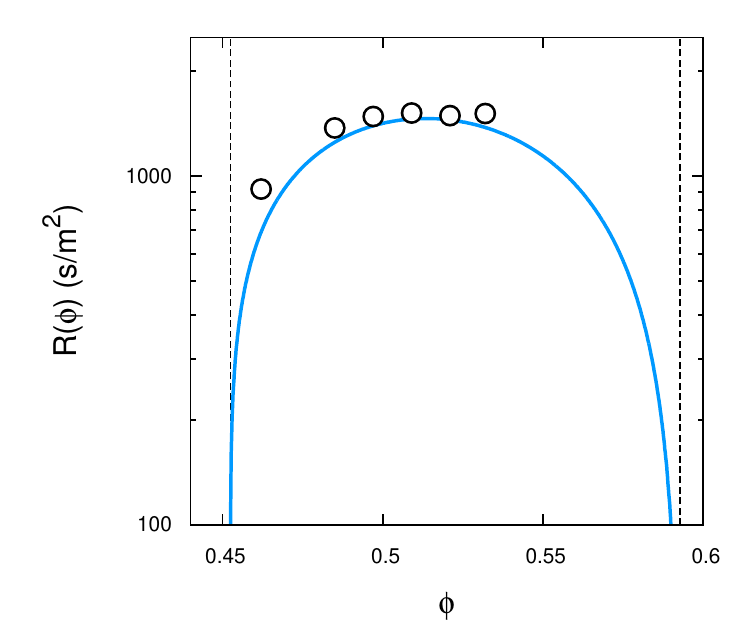}
\end{center}
\caption{\label{SR_max_Phi} Comparison of $R(\phi)$ obtained from the numerical calculation (open circles) with the prediction of Eq.~\ref{eq:SR_F} (blue line). The dashed black lines show $\phi_\text{m}$ and $\phi_0$. }
\end{figure*}

Using Eqs.~\ref{eq:F_p} and \ref{eq:DynamicEquation}, we can write out the expression for the shear rate: 
\begin{equation}
\dot{\gamma} = \left\{ \frac{\rho k^2}{\widetilde{\eta}_0} \gamma \cdot \left[ (\phi_0 - \phi_\text{m}) e^{-\gamma/\gamma^*} + \phi_\text{m} - \phi \right]^2 \right\} U_0^2, 
\label{eq:SR_F}
\end{equation} 
where we have replaced $U_\text{f}$ by $kU_0$. The maximum shear rate $\dot{\gamma}_\text{max}$ is achieved at $\gamma_\text{m}$, where the function in the curly brackets reaches its peak. By calculating the first derivative, we find that this occurs when
\begin{equation}
e^{-\gamma_\text{m}/\gamma^*} \left( 1- 2 \frac{\gamma_\text{m}}{\gamma^*} \right) = \Phi.  
\label{eq:gamma_p}
\end{equation} 
This can be evaluated numerically to find $\gamma_\text{m}$. Plugging $\gamma_\text{m}$ into Eq.~\ref{eq:SR_F}, we can see that everything in the curly brackets is independent of $U_0$. As a result, the prediction of the maximum shear rate by the model can be written as 
\begin{equation}
\dot{\gamma}_\text{max} = R(\phi) \cdot U_0^2, 
\label{eq:SR_max_CAL}
\end{equation} 
where the pre-factor $R(\phi)$ is simply a function of the packing fraction. As shown in Fig.~S\ref{SR_max_Phi}, $R(\phi)$ vanishes as $\phi \to \phi_0$ and $\phi \to \phi_\text{m}$, but in the range $\phi \in [0.462, 0.532]$, it is relatively flat. This agrees well with the numerical results shown in Fig.~2D in the main text. To extract $R(\phi)$ we fit the calculated $\dot{\gamma}_\text{max}(U_0)$ for each $\phi$ to Eq.~\ref{eq:SR_max_CAL}. The results are given by the open circles in Fig.~S\ref{SR_max_Phi}.


\begin{figure*}
\begin{center}
\includegraphics[scale=0.9]{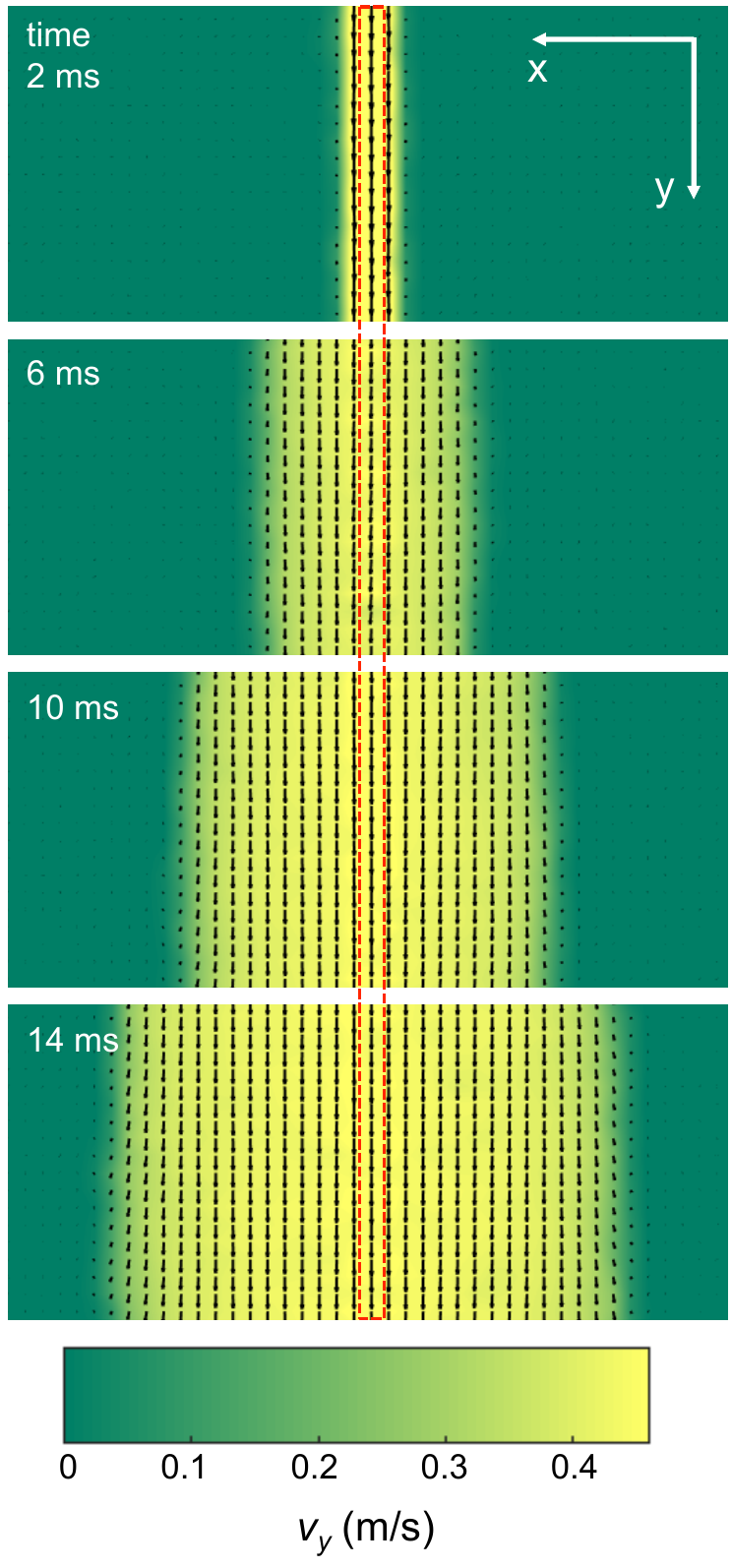}
\end{center}
\caption{\label{FlowField} Snapshots of the flow field at $\phi = 0.532$ and $U_0 = 0.46 \pm 0.02$~m/s. The plate is outlined by the red dashed line in the middle. The arrows show local velocity and the color represents the magnitude of the longitudinal component of the velocityt, $v_y$. Time of the image since the plate started to move is labeled on the upper left corner. The original video is shown in Supplementary Movie 1. }
\end{figure*}

\end{document}